\documentclass[letters]{fcs}
\usepackage{hyperref}

\title{A Design Space for Data Agents: Insights from KDD Cup 2026}
\shorttitle{A Design Space for Data Agents: Insights from KDD Cup 2026}
\hypersetup{
  pdftitle={A Design Space for Data Agents: Insights from KDD Cup 2026},
  pdfauthor={Boyan Li; Zhuowen Liang; Yupeng Xie; Xiaotian Lin; Xinyu Liu; Yizhang Zhu; Yuyu Luo},
  pdfsubject={Final manuscript, FCS-261906.R1}
}

\author[1]{Boyan~Li}
\author[1]{Zhuowen~Liang}
\author[1]{Yupeng~Xie}
\author[1]{Xiaotian~Lin}
\author[1]{Xinyu~Liu}
\author[1]{Yizhang~Zhu}
\author[1,+]{Yuyu~Luo}
\address[1]{Data Science and Analytics Thrust, Information Hub,
  The Hong Kong University of Science and Technology (Guangzhou), Guangzhou~511453,~China}
\corremail{yuyuluo@hkust-gz.edu.cn}

\fcssetup{
  received = {month dd, yyyy},
  accepted = {month dd, yyyy},
  article-id = {1},
  first-page = {1},
}

\ExplSyntaxOn
\tl_if_exist:NF \g__fcs_abstract_tl { \tl_new:N \g__fcs_abstract_tl }
\tl_if_eq:NnT \l__fcs_doi_tl { 10.1007/sxxxxx-yyy-zzzz-1 }
  { \tl_clear:N \l__fcs_doi_url_tl }
\ExplSyntaxOff

\begin{document}

\section{Introduction}
\label{sec:introduction}

Data agents use large language models (LLMs) to interpret analytical requests,
interact with data sources, and execute computational
workflows~\cite{DBLP:journals/corr/abs-2510-23587}.
Reasoning--action interaction~\cite{yao2023react} and executable code
actions~\cite{wang2024codeact} underpin these workflows.
Evidence representations shape what an agent can compute and check, while
scheduling and state management govern when that work continues and which
results can be reused. These dependencies motivate our system-level analysis.

The KDD Cup 2026 \emph{Data Agents for Complex Data Analysis}
competition~\cite{kddcup2026dataagents} offers a concrete setting for this inquiry
under common evaluation and resource constraints.
Its evaluation benchmark, \emph{DataSpace}, pairs analytical questions with
heterogeneous workspaces containing structured data, documents, and videos,
requiring agents to return complete tabular results for deterministic
evaluation~\cite{li2026dataspace}.
We analyze the final code submissions of the ten highest-ranked teams to
identify shared design choices and trade-offs across their implementations.

To organize this evidence, we develop a \emph{Data Agent Design Space} comprising
an orchestration layer for allocating and controlling work; a functional layer
for task modeling, evidence construction, candidate computation, and result
assessment; and a state layer for preserving working artifacts and their origins.
Across these ten submissions, we identify three recurring design patterns:
\textbf{(I1)} allocating finite computation according to task progress;
\textbf{(I2)} connecting computation and assessment through explicit
representations; and \textbf{(I3)} preserving reusable artifacts and their
provenance.
We use these insights to outline future directions for jointly designing and
evaluating the three layers, focusing on trade-offs in completion, correctness,
and evidence reuse.

\section{Data Agent Design Space}
\label{sec:design-space}

\begin{figure*}[t]
  \centering
  \includegraphics[width=\textwidth]{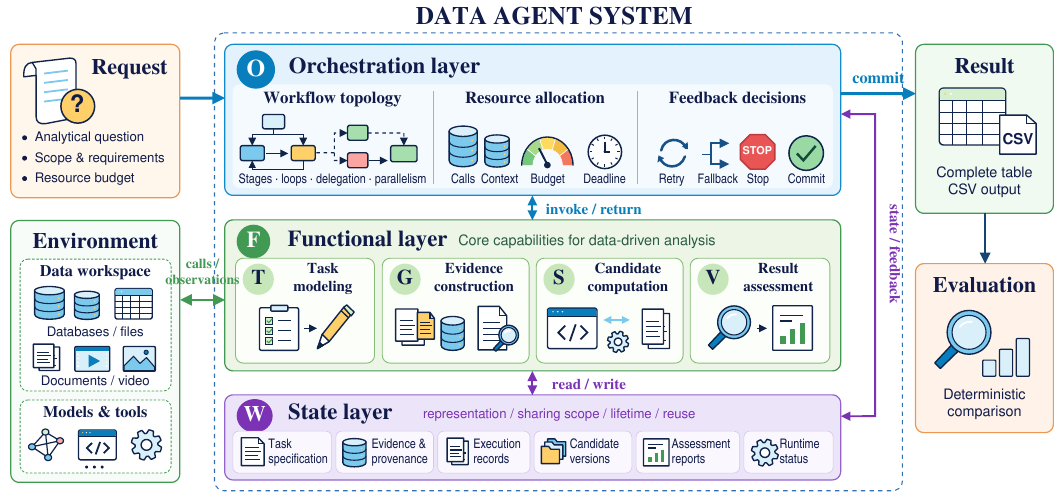}
  \caption{A three-layer design space for data agents.
  Orchestration composes four peer functions through invocation and feedback;
  working state stores their artifacts and runtime information.
  The functions access data and tools through the environment interface, while
  orchestration commits a selected candidate for external evaluation.}
  \label{fig:design-space}
\end{figure*}

\subsection{Task and System Model}
\label{sec:task-system-model}
\noindent\textbf{Task formulation.}
The task is to answer an analytical question using accessible data and bounded
resources:
\begin{equation}
\label{eq:task-formulation}
  \begin{aligned}
    \tau&=(q,R,C,B),\quad \mathcal E=(D,\mathcal U),\\
    y&=\mathcal A(\tau;\mathcal E),\quad y\models C,\quad c\preceq B.
  \end{aligned}
\end{equation}
Here $q$ is the question, $R$ identifies accessible sources in $D$, and $C$
specifies task and output requirements.
The budget $B$ bounds the run's resource-use vector $c$ componentwise, including
elapsed time and model calls.
The environment's services $\mathcal U$ support source access, model inference,
computation, and storage.
In DataSpace, $y$ is a complete table serialized as CSV for deterministic
external evaluation~\cite{li2026dataspace}.

\noindent\textbf{System formulation.}
Following Figure~\ref{fig:design-space}, we model an agent as
\begin{equation}
\label{eq:agent-system}
  \mathcal A=(O,\mathcal F,W_0,I), \qquad
  \mathcal F=\{T,G,S,V\},
\end{equation}
where $O$ orchestrates the functions in $\mathcal F$, $W_0$ is the initial
working state, and $I$ comprises the request, environment, and delivery
interfaces.
At each decision event $t$, $O$ uses the current state $W_t$ and remaining
budget $B_t$ to select function calls or commit a candidate.
For a function-invocation decision, execution proceeds as
\begin{equation}
\label{eq:agent-execution}
  \begin{aligned}
    a_t &= O(W_t,B_t),\\
    Z_t &= \operatorname{Run}(a_t,W_t;\mathcal E),\\
    W_{t+1} &= \operatorname{Update}_{O}(W_t,Z_t).
  \end{aligned}
\end{equation}
The action $a_t$ specifies functions in $\mathcal F$, input artifacts, and worker scope;
$Z_t$ contains returned artifacts and observations.
The update policy records these outputs and runtime status in working state.
Invocations may recur, run in parallel, or be delegated.
Committing a selected candidate $\hat y_t$ terminates the run through the
delivery interface $I_{\mathrm{out}}\in I$:
\begin{equation}
\label{eq:agent-commit}
  a_t=\operatorname{commit}(\hat y_t)
  \quad\Longrightarrow\quad y=I_{\mathrm{out}}(\hat y_t).
\end{equation}
\subsection{Layers and Design Choices}
\label{sec:layer-design-choices}
To derive the layers, we grouped observed mechanisms by system
responsibility. Scheduling and retries control execution,
extraction and computation transform information, and caches retain artifacts.

\noindent\textbf{Orchestration layer ($O$).}
\emph{Workflow topology} organizes calls into stages, loops, delegated work, or
parallel branches, while \emph{resource allocation} assigns model calls, context
capacity, and time to that work.
As functions return observations and assessments, \emph{feedback decisions}
determine whether to retry, fall back, stop, or commit a candidate.
Control spans source chunks, tool actions, attempts, and tasks.

\looseness=-1
\noindent\textbf{Functional layer ($\mathcal F$).}
\emph{Task modeling ($T$)} interprets the request and context into a specification
of scope, output fields, row granularity, and units, with varying degrees of
explicitness.
To supply usable tables, facts, or rules, \emph{evidence construction ($G$)}
selects, extracts, normalizes, and links source content at a chosen time and
granularity.
Using the specification and evidence, \emph{candidate computation ($S$)}
constructs and executes a solution through a chosen interface, producing a
candidate and an execution record.
\emph{Result assessment ($V$)} checks these artifacts and their supporting
specifications and evidence against applicable criteria, returning an assessment
report that informs subsequent orchestration decisions.

\noindent\textbf{State layer ($W$).}
The state layer connects invocations by preserving their artifacts and runtime
status, including specifications, evidence with provenance, execution records,
candidate versions, and assessment reports.
Their \emph{representation} in conversation context, interpreter sessions, files,
or caches determines how they can be accessed and manipulated.
\emph{Sharing scope} controls visibility across workers, attempts, or tasks,
while \emph{lifetime} controls persistence.
Policies for \emph{reuse} then determine which retained artifacts inform later
invocations, linking past work to subsequent computation and control.

\section{Design Insights}
\label{sec:insights}
To identify recurring design patterns, we compare the ten submissions
within this design space. We examine their code, configurations, and prompts,
checking ambiguous cases against the criteria in Table~\ref{tab:design-comparison}.
As competition organizers, we are authorized to analyze these
submissions for research under the competition rules~\cite{kddcup2026dataagents}.

\begin{table*}[t]
\caption{Shared design mechanisms across the ten highest-ranked final submissions.}
\label{tab:design-comparison}
\centering
\begingroup
\definecolor{insightblue}{HTML}{087FBC}
\definecolor{insightgreen}{HTML}{30753E}
\definecolor{insightpurple}{HTML}{7B32B5}
\newcommand{\present}{\ensuremath{\checkmark}}
\newcommand{\unconfirmed}{\textcolor{black!35}{\textendash}}
\fontsize{8.5pt}{10.4pt}\selectfont
\setlength{\tabcolsep}{1.5pt}
\renewcommand{\arraystretch}{0.9}
\begin{tabular}{@{}F{.055\textwidth}*{9}{F{\dimexpr(.945\textwidth-18\tabcolsep)/9\relax}}@{}}
\toprule
\textbf{Rank} &
\multicolumn{3}{c}{\textcolor{insightblue}{\textbf{Orchestration $O$}}} &
\multicolumn{3}{c}{\textcolor{insightgreen}{\textbf{Functions $\mathcal F$}}} &
\multicolumn{3}{c}{\textcolor{insightpurple}{\textbf{State $W$}}} \\
\cmidrule(lr){2-4}\cmidrule(lr){5-7}\cmidrule(lr){8-10}
& \shortstack{Task\\priority} & \shortstack{Stage\\budgets} & \shortstack{Bounded\\repair} & \shortstack{Output\\specification} & \shortstack{Structured\\evidence} & \shortstack{Evidence\\checks} & \shortstack{Cross-attempt\\reuse} & \shortstack{Source\\provenance} & \shortstack{Candidate\\retention} \\
\midrule
1 & \unconfirmed & \unconfirmed & \present & \unconfirmed & \present & \present & \present & \present & \unconfirmed \\
2 & \present & \unconfirmed & \present & \present & \present & \unconfirmed & \present & \present & \present \\
3 & \present & \unconfirmed & \present & \unconfirmed & \present & \present & \present & \present & \present \\
4 & \unconfirmed & \unconfirmed & \present & \unconfirmed & \present & \unconfirmed & \unconfirmed & \unconfirmed & \unconfirmed \\
5 & \unconfirmed & \present & \present & \unconfirmed & \present & \present & \present & \unconfirmed & \present \\
6 & \unconfirmed & \present & \present & \unconfirmed & \present & \present & \present & \unconfirmed & \present \\
7 & \present & \present & \present & \unconfirmed & \present & \present & \unconfirmed & \present & \unconfirmed \\
8 & \present & \unconfirmed & \unconfirmed & \present & \present & \unconfirmed & \present & \present & \present \\
9 & \unconfirmed & \unconfirmed & \present & \unconfirmed & \present & \present & \unconfirmed & \present & \present \\
10 & \unconfirmed & \present & \present & \present & \present & \present & \unconfirmed & \present & \unconfirmed \\
\bottomrule
\end{tabular}
\par\smallskip
\begin{minipage}{\textwidth}
\fontsize{8pt}{9.5pt}\selectfont
Ranks follow the official final leaderboard~\cite{kddcup2026dataagents}.\\
Criteria follow column order.
\textbf{Orchestration:} prioritize queued work; allocate stage-specific time/steps;
cap corrective attempts.
\textbf{Functions:} task-specific output requirements in dedicated
artifacts (code-defined or prompt-prescribed);
structured document/video evidence; checks on extracted content/coverage.
\textbf{State:} prepared evidence retained across attempts;
artifacts linked to sources or producing tools; complete candidates retained for
selection/recovery.
\end{minipage}
\endgroup
\end{table*}

\subsection{Adaptive Computation Allocation}
\label{sec:insight-orchestration}
Under a fixed budget, additional refinement competes with the
resources needed to finish. Four of the ten submissions combine stage budgets
with bounded repair (Table~\ref{tab:design-comparison}). The fifth-ranked solution
uses first-pass duration to estimate whether another pass is affordable, allowing
a 420-second margin for final selection. The runner-up's default scheduler
targets four agreeing candidates, caps attempts at ten, and limits attempts
without a valid candidate to five. These policies make continuation depend on
the progress achieved and the resources remaining.
\textbf{Computation should therefore follow observed progress while preserving
the budget needed to finish.}

\subsection{Representations as Functional Interfaces}
\label{sec:insight-functions}
\looseness=-1
Structured evidence connects semantic interpretation with
precise computation and exposes intermediate results to inspection. Seven of the
ten submissions combine structured evidence with content or coverage checks
(Table~\ref{tab:design-comparison}). The champion, for example, checks entity
coverage and row counts when assembling extracted document tables. Its submitted
table-selection record further illustrates the importance of coverage. For 33
tasks with nonempty reference sets extracted from model-generated solver
programs, mean per-task reference-table recall is 89.39\%, while the median
fraction of candidate tables selected is 6.25\%. Five tasks omit a
reference-listed table; in two cases, the omitted table receives no positive
votes across five rounds. These omissions show that agreement can coexist with
incomplete reference-table coverage. The implementation keeps unselected tables
accessible through tools, allowing source selections to be revisited.
\textbf{A representation shapes both what the agent can compute and which errors
it can detect.}

\subsection{Artifact Reuse and Provenance}
\label{sec:insight-state}
Six of the ten submissions retain prepared evidence across
attempts, and four of these also preserve provenance
(Table~\ref{tab:design-comparison}). The runner-up reuses extracted document tables
across candidate attempts, avoiding repeated source preparation. The
seventh-ranked solution records the producing tool and its arguments so that
the complete result can be regenerated at submission. Such records preserve how
an artifact was obtained, making its derivation available for later inspection.
When candidates reuse the same evidence, they also inherit any errors it
contains. Candidate agreement must therefore be interpreted in light of shared
evidence dependencies.
\textbf{State should preserve both reusable work and its provenance, so that
additional computation can be distinguished from additional evidence.}

\section{Discussion and Future Opportunities}
\label{sec:discussion}
\noindent\textbf{(O1)} The three insights motivate joint design of the layers around the value of
additional computation.
Future agents could use evidence dependencies recorded in working state,
together with assessment findings, to guide resource allocation.
When a suspected error originates in evidence shared by several candidates,
the agent could revisit the source before generating further candidates.
If the evidence changes, recorded dependencies could identify which downstream
results require recomputation.
This would connect the granularity of evidence construction to the scope of
repair, enabling the agent to address the underlying problem while preserving
unaffected work.

\looseness=-1
\noindent\textbf{(O2)} These observed patterns motivate controlled comparisons
that hold the model, task set, and total budget fixed while jointly varying
orchestration policies and evidence sharing.
Injecting known errors into extracted tables would then test whether agents
can locate the faulty evidence and repair affected answers.
Alongside final correctness, evaluation should measure task completion and
the resources consumed by diagnosis and recovery.
Such comparisons would clarify when reuse supports adaptive allocation and
when revisiting shared evidence is necessary.

\fontsize{8.5pt}{10pt}\selectfont
\section*{Acknowledgements}
We thank the participating teams of the KDD Cup 2026 Data Agents for Complex
Data Analysis competition for their submissions, which provided the basis for
this study.
This work was supported by the NSF of China (62402409);
Youth S\&T Talent Support Programme of Guangdong Provincial Association for
Science and Technology (SKXRC2025461);
the Young Talent Support Project of Guangzhou Association for Science and
Technology (QT-2025-001);
Guangzhou Basic and Applied Basic Research Foundation
(2026A1515010269, 2025A04J3935, 2023A1515110545);
Guangzhou-HKUST(GZ) Joint Funding Program (2025A03J3714);
and Research and Development of Heterogeneous Computing Interconnection and
Scheduling Software Stack (YF202400000003).

\section*{Competing interests}
All authors participated in organizing the KDD Cup 2026 Data Agents for Complex
Data Analysis competition. Boyan Li and Yuyu Luo served as General Chairs, and
the other authors served as Committee Chairs.

\bibliographystyle{fcs}
\bibliography{ref}

\begin{thebibliography}{1}

\bibitem{DBLP:journals/corr/abs-2510-23587}
Zhu Y, Wang L, Yang C, Lin X, Li~B, Zhou W, Liu X, Peng Z, Luo T, Li~Y, Chai C,
  Chen C, Di~S, Fan J, Sun J, Tang N, Tsung F, Wang J, Wu~C, Xu~Y, Zhang S,
  Zhang Y, Zhou X, Li~G, Luo Y.
\newblock A survey of data agents: Emerging paradigm or overstated hype?
\newblock CoRR, 2025, abs/2510.23587

\bibitem{yao2023react}
Yao S, Zhao J, Yu~D, Du~N, Shafran I, Narasimhan K, Cao Y.
\newblock {ReAct}: Synergizing reasoning and acting in language models.
\newblock In: Proceedings of the Eleventh International Conference on Learning
  Representations.
\newblock 2023

\bibitem{wang2024codeact}
Wang X, Chen Y, Yuan L, Zhang Y, Li~Y, Peng H, Ji~H.
\newblock Executable code actions elicit better {LLM} agents.
\newblock In: Proceedings of the 41st International Conference on Machine
  Learning.
\newblock 2024,  50208--50232

\bibitem{kddcup2026dataagents}
{KDD Cup 2026} .
\newblock {KDD Cup 2026}: Data agents for complex data analysis, 2026.
\newblock Accessed September 16, 2026

\bibitem{li2026dataspace}
Li~B, Liang Z, Xie Y, Lin X, Luo T, Liu X, Zhu Y, Peng Z, Li~Y, Zhang Z, Zhang
  J, Tang N, Li~G, Luo Y.
\newblock {DataSpace}: Benchmarking data agents for verifiable analytics over
  heterogeneous workspaces.
\newblock arXiv preprint arXiv:2608.03451, 2026

\end{thebibliography}

\end{document}